\documentclass[conference,a4paper]{IEEEtran}
\usepackage[left=1.43cm,right=1.43cm,top=1.82cm,bottom=4.3cm]{geometry}

\usepackage[algo2e]{algorithm2e} 
\usepackage{amsmath,amssymb,amsmath,amsfonts}
\usepackage{bm}
\usepackage{bbm}
\usepackage{graphicx}
\usepackage{cite}
\usepackage{epstopdf}
\usepackage{gensymb}
\usepackage{color}
\usepackage{lipsum}
\usepackage{float}
\usepackage{epstopdf}
\usepackage[mathcal]{eucal}
\usepackage{subfiles}
\usepackage[T1]{fontenc}
\usepackage{siunitx}
\usepackage{tabulary}
\usepackage{array}
\usepackage{adjustbox}
\usepackage[dvipsnames]{xcolor}
\usepackage{cite}
\usepackage{times}
\usepackage{placeins}
\usepackage{textcomp}
\usepackage[font=small]{caption}
\usepackage{flushend}
\usepackage{color, colortbl}
\usepackage{tabularx}
\usepackage{bm}
\usepackage{enumerate}
\usepackage{tikz}
\usepackage{pgfplots}
\usepackage{epstopdf}
\usetikzlibrary{shapes,arrows}
\usepackage[utf8]{inputenc}
\usepackage{mathtools}
\usepackage[utf8]{inputenc}
\usepackage{xcolor}
\usepackage{tikz}
\usetikzlibrary{chains,arrows,calc,positioning}
\usepackage{amssymb}% http://ctan.org/pkg/amssymb
\usepackage{pifont}% http://ctan.org/pkg/pifont
\usepackage{soul}
\usepackage{breqn} % to split equations using dmath env
\usepackage{booktabs} % nice rules in tables
\usepackage{multirow}
\usepackage{enumitem}
\usepackage{tabulary}
\usepackage[normalem]{ulem}
\usepackage{dblfloatfix}
\usepackage{makecell}
\usepackage{lipsum}
\usepackage{amsthm}
\usepackage{comment}
\usepackage{filecontents}
\usepackage{subcaption} 
\usepackage[nolist,nohyperlinks]{acronym}

\newtheoremstyle{mystyle}%                % Name
  {}%                                     % Space above
  {}%                                     % Space below
  {\itshape}%                             % Body font
  {}%                                     % Indent amount
  {\bfseries}%                            % Theorem head font
  {.}%                                    % Punctuation after theorem head
  { }%                                    % Space after theorem head, ' ', or \newline
  {}%                                     % Theorem head spec (can be left empty, meaning `normal')
\theoremstyle{mystyle}

\newlength \figwidth
\definecolor{bittersweet}{rgb}{1.0, 0.44, 0.37}
\definecolor{glaucous}{rgb}{0.38, 0.51, 0.71}
\definecolor{gainsboro}{rgb}{0.86, 0.86, 0.86}
\definecolor{babyblueeyes}{rgb}{0.63, 0.79, 0.95}
\definecolor{silver}{rgb}{0.75, 0.75, 0.75}
\definecolor{neoncarrot}{rgb}{1.0, 0.64, 0.26}
\definecolor{Gray}{gray}{0.9}
\definecolor{LightCyan}{rgb}{0.88,1,1}
\definecolor{BackgroundLightBlue}{rgb}{0.97,0.97,1}
\definecolor{BackgroundGray}{gray}{0.98}

\makeatletter

 \let\oldforeign@language\foreign@language
 \DeclareRobustCommand{\foreign@language}[1]{%
   \lowercase{\oldforeign@language{#1}}}

\def\nb0{{\mathbf{0}}}
\def\nb1{{\mathbf{1}}}

\makeatother

\IEEEoverridecommandlockouts
\IEEEoverridecommandlockouts\IEEEpubid{\makebox[\columnwidth]{ \hfill} \hspace{\columnsep}\makebox[\columnwidth]{ }}
\begin{document}

% This code is to reduce the list of authors by using et. al:
\bstctlcite{IEEEexample:BSTcontrol}

\title{Capacity and Energy Trade-Offs in FR3 6G Networks Using Real Deployment Data\vspace{-0.5cm}}

\author{\IEEEauthorblockN{
David L\'{o}pez-P\'{e}rez$^{\sharp}$,\vspace{0.1cm} 
Nicola Piovesan$^{\flat}$,\vspace{0.1cm} %
% , and Giovanni Geraci$^{\dagger\,\star}$
and Matteo Bernabè$^{\sharp}$
}
\\ \vspace{-0.4cm}
\normalsize\IEEEauthorblockA{
\makebox[0.5\textwidth][r]{$^{\sharp}$\emph{Universitat Politècnica de València, Spain}} \enspace \enspace
\makebox[0.5\textwidth][l]{$^{\flat}$\emph{Huawei Technologies, France}} 
%\makebox[0.5\textwidth][r]{$^{\star}$\emph{Universitat Pompeu Fabra, Barcelona, Spain}} \enspace \enspace
%\makebox[0.5\textwidth][l]{$^{\dagger}$\emph{Telefónica Scientific Research, Spain}}
\vspace{-0.8cm}
}
%\thanks{This research was supported by the Generalitat Valenciana, Spain, through the  CIDEGENT PlaGenT,  Grant CIDEXG/2022/17, Project iTENTE, and the action CNS2023-144333, financed by MCIN/AEI/10.13039/501100011033 and the European Union “NextGenerationEU”/PRTR.}
}

\maketitle
%\IEEEpeerreviewmaketitle

% \vspace{-0.3cm}
\begin{abstract}

This article presents a data-driven system-level analysis of multi-layer 6G networks operating in the upper mid-band (FR3: 7–24\,GHz). 
Unlike most prior studies based on \ac{3GPP} templates, 
we leverage real-world deployment and traffic data from a commercial 4G/5G network in China to evaluate practical 6G strategies.
Using Giulia---a deployment-informed system-level heterogeneous network model---we show that 6G can boost median throughput by up to 9.5× over heterogeneous 4G+5G deployments, 
but also increases power usage by up to 59\,\%. 
Critically, co-locating 6G with existing sites delivers limited gains while incurring high energy cost. 
In contrast, non-co-located, traffic-aware deployments achieve superior throughput-to-watt efficiency,
highlighting the need for strategic, \ac{UE} hotspot-focused 6G planning.
\end{abstract}
%\begin{IEEEkeywords}
%\end{IEEEkeywords}
\acresetall

\vspace{-0.0cm}
\section{Introduction}

In contrast to the ambitious roadmaps often envisioned for the \ac{6G} of the mobile network technology, 
we argue that its commercial realization will likely center on a narrower set of practical features. 
Much like \ac{5G} deployments, 
which ultimately realized a few high-impact technologies---most notably \ac{mMIMO}, \ac{URLLC}, and network slicing~\cite{dahlman2023}---,
we expect \ac{6G} to follow a similarly selective trajectory. 
This pragmatic focus is especially relevant in the current macroeconomic climate, 
where rising infrastructure costs and uncertain revenue models are reshaping investment strategies across the telecommunications sector.

Among the features likely to define early \ac{6G} rollouts, 
we identify three foundational enablers that are likely to be both technically mature and economically viable by 2030~\cite{LopezPerez2025}:
\begin{itemize}
    \item 
    \emph{Extreme \ac{mMIMO} at \ac{cm}-wave (7--24~GHz):} 
    A cornerstone for scalable, high-capacity deployments,
    \ac{cm}-wave extreme \ac{mMIMO} will combine wide bandwidth availability with favorable propagation and implementation characteristics,
    making it ideal for next-generation urban and indoor networks.

    \item \emph{\Ac{AI}:}
    Deeply embedded across the protocol stack, 
    \ac{AI} will allow the smart and dynamic orchestration of heterogeneous 4G/\ac{5G}/\ac{6G} networks, 
    supporting functions such as load balancing, interference mitigation, and energy-efficient scheduling.

    \item \emph{Sensing:}
    By equipping networks with environmental awareness,
    sensing technologies will enable new services such as autonomous driving and smart infrastructure.
    Additionally, sensing data will enhance \ac{AI}-driven network intelligence through real-time contextual feedback.
\end{itemize}

Among these, 
\ac{cm}-wave extreme \ac{mMIMO} stands out as the anchor technology for delivering the multi-gigabit-per-second capacity demanded by emerging \ac{6G} use cases.
It can achieve this capacity,
while circumventing many of the deployment challenges inherent to \ac{mm}-wave systems, 
including high energy usage, blockage sensitivity, as well as poor indoor penetration and challenging mobility management~\cite{BjoKarKol2025}.

\vspace{-0.0cm}
\subsection{Network Optimization Challenge}

Designing and optimizing the resulting complex, heterogeneous 4G/\ac{5G}/\ac{6G} networks, 
especially in dense urban environments,
will increasingly exceed the limits of current human-based engineering. 
This growing complexity calls for a shift toward AI-driven orchestration and adaptation. 
However, the effectiveness of these \ac{AI}-based strategies critically depends on the availability of accurate, realistic models for training and evaluation.%~\cite{}.

Yet much of the current cellular modeling landscape remains rooted in \ac{3GPP}-style deployment templates and abstract traffic assumptions~\cite{3GPP38901},
which fail to capture the local spatial, temporal, and topological complexities of real-world networks. 
This gap can lead to misleading assessments of system performance, 
especially in multi-layer deployments that integrate legacy and next-generation technologies.

\emph{What is needed instead is a data-driven, deployment-informed modeling framework}---one that reflects actual site layouts, hardware configurations, user behaviors, and traffic patterns. 
Such models are essential for evaluating true system-level trade-offs in coverage, throughput, latency, and energy usage under realistic operational conditions.

\vspace{-0.0cm}
\subsection{Contributions of This Work}

In this article, 
we present a comprehensive, data-driven system model for---and analysis of---multi-layer \ac{6G} deployments operating in the upper mid-band spectrum,
building upon and significantly advancing our earlier work~\cite{LopezPerez2025}. 
This study departs from synthetic assumptions and embraces realism through direct integration of live network data,
enabling a much more accurate exploration of next-generation architectures.

At the heart of our analysis lies \texttt{Giulia}---a high-fidelity, system-level simulator purpose-built for evaluating heterogeneous cellular networks.
Unlike conventional tools that rely primarily on abstracted or idealized models,
\texttt{Giulia} is designed to integrate real-world engineering data and \ac{AI} models whenever available. 
In scenarios where such data is not accessible,
it remains compatible with standard \ac{3GPP} assumptions,
ensuring analytical continuity. 
The digital twin captures detailed aspects of \ac{BS} and \ac{UE}, network configuration, radio propagation, \ac{mMIMO}, scheduling behavior, beamforming strategies, and power consumption with deployment-grade precision.
%revise las tsentence : At the heart of our analysis lies \texttt{Giulia}
%---a high-fidelity, system-level simulator purpose-built for the evaluation of heterogeneous cellular networks.
%Unlike conventional tools that rely primarily on abstracted or idealized models,
%\texttt{Giulia} is designed to integrate real-world engineering data and \ac{AI} models whenever available. 
%In scenarios where such data is not accessible,
%it remains compatible with standard \ac{3GPP} assumptions,
%ensuring analytical continuity. 
%The digital twin captures detailed aspects of \ac{BS} and \ac{UE}, network configuration, radio propagation, \ac{mMIMO}, scheduling behavior, beamforming strategies, and power consumption with deployment-grade precision.
Importantly, and unlike most tools, \texttt{Giulia} supports per-cell radio customization---each cell can be powered by radios with distinct characteristics and configurations.

Key contributions of this article include:
\begin{itemize}
    \item \textit{Deployment-Informed Network Modeling:} 
    Leveraging proprietary datasets from a commercial urban network in China,
    we extract true-to-life spatial layouts, engineering parameters, and load distributions across 4G, \ac{5G}, and synthetic \ac{6G} layers.

    \item \textit{Realistic System-Level Simulation:} 
    Using \texttt{Giulia},
    we simulate end-to-end system behavior with fidelity using a mix of expert-, \ac{3GPP}-, and \ac{AI}-based models,
    offering a holistic view of capacity and energy trade-offs.

    \item \textit{Comparative Evaluation of Deployment Strategies:} 
    We examine multiple architectural variants---including co-located, non-co-located, and traffic-driven hotspot deployments of \ac{6G} \ac{cm}-wave cells---and benchmark them against legacy 4G/\ac{5G} baselines to uncover insights into spectrum utilization, energy efficiency, and spatial densification.
\end{itemize}

\begin{comment}
Our results reveal that non-co-located \ac{6G} deployments
---strategically placed at high-traffic hotspots---
can achieve over $15\times$ improvements in \ac{UE} throughput relative to \ac{4G} baselines, 
albeit with moderate increases in energy usage.
While co-locating \ac{6G} with existing \ac{5G} sites yields more modest performance gains, 
it offers cost and energy efficiency benefits.
These findings highlight the trade-offs between capacity, energy, and deployment complexity, 
underscoring the importance of context-aware network deployment decisions.
Data-grounded, scenario-specific modeling thus emerges as a critical enabler for guiding strategic choices in the design and evolution of \ac{6G} networks.
%David: To be rewritten based on the results. 
\end{comment}

The remainder of this article is organized as follows. 
Section~\ref{sec:extremeMIMO} reviews recent advances in \ac{cm}-wave extreme \ac{mMIMO}. 
Section~\ref{sec:system_model} introduces \texttt{Giulia}, 
our data-driven system model. 
%Section~\ref{sec:deploy_eval} outlines the deployment configurations and \acp{KPI} analyzed in this study. 
Section~\ref{sec:results} presents our performance evaluation of heterogeneous 4G/\ac{5G}/\ac{6G} networks, 
highlighting trade-offs across different deployment strategies. 
Finally, Section~\ref{sec:conclusions} concludes the article with a summary of findings and future research directions.

\vspace{-0.0cm}
\section{FR3 cm-Wave Extreme MIMO}
\label{sec:extremeMIMO}

The upper mid-band spectrum from 7 to 24~GHz, 
referred to by \ac{3GPP} as \ac{FR3}~\cite{3GPP38820}, 
is a strong candidate for early \ac{6G} deployments. 
\ac{FR3} sits between sub-6~GHz spectrum (FR1), 
which offers wide-area coverage but limited bandwidth, 
and \ac{mm}-wave bands FR2, 
which offer very large bandwidth but suffer from severe propagation losses. 
This intermediate position makes \ac{FR3} attractive for dense urban networks that need multi-gigabit capacity without the fragility typically associated with FR2.

To translate \ac{FR3} bandwidth into cell-level capacity, 
\acp{BS} must also deliver substantially higher beamforming and spatial-multiplexing gains than today’s sub-6~GHz systems. 
This requires antenna-array and radio designs that scale to very large active panels while keeping cost and power consumption manageable.
This trend is driving \emph{extreme} \ac{mMIMO},
with thousands of antenna elements and well over 128 TRX chains,
enabling narrow beams and high-order spatial multiplexing~\cite{Nokia2024cmWave,Huawei2024cmWave}.
When paired with \ac{UE} devices equipped with up to 64 antennas, 
FR3-based extreme mMIMO may support up to 20 spatial streams per device, 
enabling peak data rates approaching 50~Gbps~\cite{Huawei2024cmWave}.

\subsection{Opportunities and Performance Advantages}

Compared to FR2 deployments:
\begin{itemize}
\item 
    \ac{FR3} provides channel bandwidths comparable to FR2 (up to 400~MHz) while offering more robust coverage in dense urban layouts, enabling broader service areas and more reliable \ac{NLoS} connectivity.
\item
    For outdoor-to-indoor users in particular, the longer wavelength reduces penetration loss and sensitivity to shadowing, improving cell-edge stability and indoor service continuity~\cite{Kang2024}.
\item 
    Beam training is less complex due to wider beams and slower beam/angle variation under mobility, 
    reducing control overhead.
\item
    Lower carrier frequencies improve hardware efficiency,
    with less stringent phase-noise and power-amplifier requirements~\cite{3GPP38820}.
\end{itemize}

\Ac{FR3} thus delivers much of the spectral capacity of FR2 while avoiding critical deployment drawbacks such as poor mobility management, bad indoor penetration, blockage sensitivity, and excessive power demand~\cite{BjoKarKol2025}.

\vspace{-0.0cm}
\subsection{Deployment Challenges at FR3}

While \ac{FR3} presents a compelling balance between capacity and coverage, 
its practical realization demands navigating a range of physical, architectural, and regulatory hurdles. 
Chief among these are channel propagation losses and the scaling of energy usage-issues that, although less severe than in FR2 (\ac{mm}-Wave),
remain significant bottlenecks.
These challenges cluster into five primary domains:

\subsubsection{Propagation and Coverage Trade-offs}

The key propagation challenge of \ac{FR3} lies in its frequency-dependent path loss.
Empirical measurements indicate that moving from 3.5\,GHz to 10\,GHz induces up to a 10\,dB degradation under both \ac{LoS} and \ac{NLoS} conditions~\cite{3GPP38901,Huawei2024cmWave}. 
Above 17\,GHz, 
this loss profile steepens further, 
severely constraining link margins and coverage depth.

To overcome this gap, 
\ac{FR3} networks must rely on compensatory mechanisms such as high-gain directional beamforming---which requires more complex antenna arrays and precise beam control than in FR1---and/or denser network architectures---including microcells, traffic-driven hotspots, and elevation-aware placements.
These approaches are essential to maintain target throughput and latency,
particularly in urban environments where propagation is highly variable.
These dense and directional deployments, however, increase hardware complexity and energy demand, making power efficiency a cross-cutting design priority in future \ac{FR3} systems.

\subsubsection{Beamforming and Antenna Design Trade-offs}

Bridging the $\sim$10\,dB path loss at \ac{FR3} necessitates highly directional beams,
placing the design burden on practical beamforming realizations and their cost–power trade-offs.
Fully digital arrays deliver the greatest spatial flexibility and multiuser capacity,
but their one-\ac{TRX}-per-element structure drives prohibitive power and data-conversion overheads as array sizes and bandwidths grow.
Analog-dominant designs simplify hardware and reduce energy use,
yet restrict dynamic beam control and wideband operation~\cite{dahlman2023}.
Hybrid beamforming---where antenna elements are combined into fixed \ac{RF} subarrays of three to six elements, 
each driven by a \ac{TRX}---offers a practical middle ground and is prevalent today.
For \ac{FR3},
its configuration must be re-optimized to handle wideband beam-squint, calibration of large apertures, and scalability in cost and power.
The final design balance depends on subarray topology, phase-shifter or true-time-delay networks, and the desired level of digital precoding flexibility, all aiming to enable scalable multi-layer deployments.

\subsubsection{Energy Efficiency and Thermal Constraints}

Building on these architectural considerations,
energy usage remains one of the most critical bottlenecks for extreme \ac{mMIMO} in \ac{FR3}.
Current \ac{5G} radios already exhibit high power draw, 
even in idle mode, 
due to the static energy drain of power amplifiers and always-on \ac{TRX} chains. 
A typical configuration with 64 \ac{TRX} chains and 192 antenna elements consumes over 250\,W in idle,
with each \ac{TRX} contributing 2–3\,W 
regardless of traffic load~\cite{Nokia2023extremeMassiveMIMO}.

If \ac{6G} systems scale along the same architectural trajectory---toward 128–256 \ac{TRX} chains and significantly wider bandwidths---, per-radio energy usage at low loads could easily exceed 1~kW~\cite{Huawei2024cmWave}.
More efficient hardware is essential, 
yet insufficient on its own.
Equally critical is the aggressive use of power-saving mechanisms, 
including \ac{TRX} sleep modes and carrier shutdown schemes~\cite{LopDomPio2021}. 
Today, these features remain underutilized in live networks, 
primarily due to operator concerns over coverage degradation and service reliability.

\begin{comment}
\subsubsection{RF Front-End and Hardware Limitations}

\ac{FR3} operation places notable strain on \ac{RF} front-end components, 
particularly above 12\,GHz~\cite{3GPP38820}. 
Power amplifiers must meet strict linearity and spectral mask requirements, 
requiring high back-off and reducing efficiency—especially under high-\ac{PAPR} waveforms like \ac{OFDM}.
These non-linearities also increase intermodulation and in-band distortion.
\ac{CMOS} and \ac{SiGe} technologies struggle to deliver sufficient output power beyond 10\,GHz, 
while \ac{GaN} offers higher performance at the expense of cost and thermal complexity.
% [Matteo] Added missing acronyms. \
% [Matteo] Overall, this paragraph above is not totally clear to me. In particular, after "These non-linearities ..."

Filters and duplexers degrade with frequency, 
exhibiting higher insertion loss and limited selectivity, worsening noise figures in interference-heavy environments. 
Phase noise also increases, 
restricting modulation fidelity unless mitigated by complex LO architectures.
% [Matteo] I do not know what is LO here

These constraints suggest that the lower portion of \ac{FR3} (7–12\,GHz) offers a more practical balance between bandwidth, hardware feasibility, and integration efficiency.
\end{comment}

\subsubsection{Regulatory Fragmentation and Coexistence Requirements}

The 7--24\,GHz spectrum remains fragmented across regions,
often shared with radar, satellite, and Earth observation services~\cite{3GPP38820}. 
For example, 7–8.5\,GHz overlaps with radar, 
10.7–12.7\,GHz with satellite downlinks, 
and 13.75–14.5\,GHz with uplink Earth stations. 
Coexistence with passive incumbents such as radio astronomy and \ac{EESS} requires mitigation strategies like geo-fencing and beam nulling.
The pace and scope of international harmonization will directly shape vendor roadmaps and deployment timelines.

\vspace{-0.0cm}
\section{System Model for Data-Driven Multi-Layer 6G Networks}
\label{sec:system_model}

To explore future deployment strategies and assess their implications on performance and energy usage,
we develop a model grounded in a real-world commercial deployment in a dense urban area in China, covering 6.2\,km$^2$.
The model incorporates empirical data from legacy \ac{4G} \ac{LTE} and operational \ac{5G} \ac{NR} networks, 
including actual \ac{BS} locations, operating frequencies, bandwidth allocations, engineering parameters, radio configurations, and traffic load distributions.

Building on this foundation, 
a synthetic \ac{6G} layer is overlaid to simulate prospective deployment scenarios.
All simulations are conducted using \texttt{Giulia}, 
which accurately models the behavior of heterogeneous networks using a mix of expert-, \ac{3GPP}-, and \ac{AI}-based models. 

%%%%%%%%%%%%%%%%%%%%%%%%%%%%%%%%%%%%%%%%%%%%%%%%%%%%%%%%%%%%%%%%%
\vspace{-0.0cm}
\subsection{Multi-Layer Network Deployment}

The network comprises three distinct layers:

\subsubsection{4G \ac{LTE} Layer (Empirical)}

The \ac{4G} layer is modeled using data extracted directly from the commercial network. 
It consists of 47 physical sites, each typically configured in a tri-sector layout, 
amounting to a total of 204 \ac{LTE} cells. 
These cells operate across five frequency bands: 
n3 (1.815\,GHz), n34 (2.02\,GHz), n39 (1.89\,GHz), n40 (2.33\,GHz), and n41 (2.62\,GHz). 
This layer exemplifies the heterogeneity of real-world macro-cell deployments in both spectrum use and hardware configuration:

\begin{itemize}
    \item 
    184 cells operate with 20\,MHz bandwidth, while 20 cells use 10\,MHz.
    \item 
    Transmit power ranges from 40 to 52\,dBm, with a median of 45.9\,dBm.
    \item 
    The number of \acp{TRX} per cell spans from 2 to 64, with a median of 4.
    A small subset of \ac{LTE} cells equipped with 64 \acp{TRX} implements \ac{mMIMO} using planar arrays.
\end{itemize}

\begin{comment}
These empirically derived configurations highlight the diversity in operational settings
---far from the uniformity assumed in typical \ac{3GPP} models where all cells share identical parameters.  
\end{comment}

\subsubsection{5G \ac{NR} Layer (Empirical)}

The \ac{5G} layer is also modeled based on the empirical deployment data. 
It consists of 45 \ac{NR} cells distributed across 15 sites 
% non-co-located with the \ac{4G} sites, 
non-co-located with the \ac{4G} ones,
each operating in band n41 at 2.703\,GHz. 
Unlike the \ac{4G} layer, 
the \ac{5G} deployment reflects a more uniform configuration typical of commercial \ac{mMIMO} rollouts: 
fixed 100\,MHz bandwidth,
with 64 \acp{TRX} per cell,
and a transmit power ranging from 51.3 to 54.7\,dBm, with a mean of 53.2\,dBm.

\begin{comment}
\begin{itemize}
    \item Fixed 100\,MHz bandwidth,
    \item 64 \acp{TRX} per cell,
    \item Transmit power ranging from 51.3 to 54.7\,dBm, with a mean of 53.2\,dBm.
\end{itemize}
\end{comment}

\begin{comment}
The model resembles real-world installations in dense urban regions and serves as a reliable reference for evaluating dual-layer \ac{LTE}/\ac{NR} architectures.     
\end{comment}

\subsubsection{6G Layer (Synthetic)}

To explore future deployment strategies,
a third layer representing \ac{6G} is \emph{synthetically added}. 
This layer consists of another 45 cells operating at 10\,GHz (within the \ac{FR3} band), 
following two distinct placement strategies:
\begin{itemize}
    \item 
    \emph{Co-located micro cells:}
    Deployed at the same sites as \ac{5G} \acp{BS}, using high-power (55\,dBm) configurations.
    \item \emph{Non-co-located cells:} 
    Independently placed at traffic hotspots, using either micro (55\,dBm) or pico (52\,dBm) radios.
\end{itemize}
All \ac{6G} radios are assumed to operate with 200 or 400\,MHz bandwidth and employ 128 or 256\,\acp{TRX}, 
enabling advanced spatial multiplexing and highly directional beamforming. 

%%%%%%%%%%%%%%%%%%%%%%%%%%%%%%%%%%%%%%%%%%%%%%%%%%%%%%%%%%%%%%%%%
\vspace{-0.0cm}
\subsection{User Equipment Distribution (Calibrated from Data)}

The spatial distribution of \acp{UE} is modeled using a hybrid, data-calibrated approach. 
Although exact \ac{UE} coordinates are unavailable, 
hourly per-cell \ac{UE} count statistics extracted from live commercial network data provide a robust foundation for realistic traffic modeling.

Focusing on a representative peak-hour snapshot,
the number of active \acp{UE} per cell is derived directly from these measurements,
and uniformly distributed within the corresponding estimated coverage area. 
The resulting baseline scenario includes 3604 active \acp{UE} across the analyzed urban region.

To project the scenario toward future 6G traffic conditions, 
the baseline is augmented with 15 synthetic traffic hotspots, 
each randomly located within the most heavily loaded cells. 
Each hotspot hosts 40 \acp{UE}, 
randomly scattered within a 40\,m radius, 
with a minimum inter-hotspot separation of 80\,m to avoid excessive spatial interference coupling.

Finally, each \ac{UE} is independently classified as indoor or outdoor 
with an 80\% probability of being indoors, 
reflecting observed user behavior in dense urban environments.

%%%%%%%%%%%%%%%%%%%%%%%%%%%%%%%%%%%%%%%%%%%%%%%%%%%%%%%%%%%%%%%%%
\vspace{-0.0cm}
\subsection{Antenna Configurations}

\Ac{BS} antenna arrays are inferred from the number of \acp{TRX} per cell and calibrated based on known commercial hardware practices:
\begin{itemize}
    \item 
    \emph{4G:} 
    Cells are equipped with vertically oriented, cross-polarized linear arrays. 
    Most radios use compact column configurations (e.g., 2×1, 4×1, or 8×1), 
    while a limited subset of high-capacity sites with 64 \acp{TRX} employ 8×4 planar arrays.
    \item 
    \emph{5G:} 
    All cells employ 8×4 cross-polarized planar arrays.
    \item 
    \emph{6G:} 
    Synthetic cells are assumed to use 16×4 or 32x4 cross-polarized planar arrays,
    depending on whether they have 128 or 256 \acp{TRX}.
    Future cells are modeled with larger planar apertures---either $16\times16$ or $32\times16$ cross–polarized arrays---paired with $128$ or $256$ \acp{TRX}, respectively.
\end{itemize}
Each antenna element in all arrays is modeled as in~\cite{3GPP38901}.

\Acp{UE} are assumed to be equipped with two cross-polarized antennas, 
supporting dual-polarization reception for downlink spatial multiplexing.

%%%%%%%%%%%%%%%%%%%%%%%%%%%%%%%%%%%%%%%%%%%%%%%%%%%%%%%%%%%%%%%%%
\vspace{-0.0cm}
\subsection{Propagation}

The propagation model follows the \ac{3GPP} \ac{UMa} and \ac{UMi} standard recommendations \cite{3GPP38901}, 
except for the fast fading component, 
which is modeled using a Rician distribution,
whose K-factor is consistent with such \ac{3GPP} guidelines. 
\acp{BS},
whose antenna arrays, 
are placed at heights greater than 15\,m are categorized as \ac{UMa},
while those below this threshold are treated as \ac{UMi}.

%%%%%%%%%%%%%%%%%%%%%%%%%%%%%%%%%%%%%%%%%%%%%%%%%%%%%%%%%%%%%%%%%
\vspace{-0.0cm}
\subsection{SSB Beam Configuration and Initial Access}

Unlike legacy \ac{4G} systems that rely on wide-angle, sectorized \acp{CRS}, 
both \ac{5G} and \ac{6G} define initial access through directional \acp{SSB}~\cite{dahlman2023}. 
Here, 
% [Matteo]: I added "here", as SSB can be potentially trasnmitted in any desired direction
% Note that it is totally reasonable using DFT here. As the direction of SSB are not know, as they are 'sensitive' info.
each \ac{SSB} is beamformed using a codeword from a 2D-\ac{DFT} codebook 
and transmitted through a single polarization panel. 
Crucially, these beams are transmitted sequentially in a time-multiplexed fashion, 
allowing \acp{UE} to detect and evaluate them one at a time. 
This approach enhances spatial selectivity during initial access,
improving coverage and link robustness.

The number of \ac{SSB} beams per cell increases with each generation to enable finer spatial resolution:
\begin{itemize}
    \item \emph{5G:} All cells use 8 SSB beams.
    \item \emph{6G:} Synthetic cells are assumed to use 16 or 32 SSB beams, depending on whether using $128$ or $256$ \acp{TRX}~\cite{LopezPerez2025}.
\end{itemize}

During association, 
each \ac{UE} measures the \ac{RSRP} of all detected \ac{SSB} beams 
and selects the strongest one as its serving beam and corresponding cell. 
If multiple technologies are available, 
a priority-based reselection mechanism is applied. 
Each \ac{UE} monitors the \ac{RSRP} of available \ac{SSB} beams across all technologies, 
and may reselect to a higher-generation layer if the received signal exceeds a predefined threshold: 
$-110$\,dBm for \ac{5G} and $-108$\,dBm for \ac{6G}. 
This ensures that transitions occur only when higher-generation coverage is strong enough to sustain quality service. 
We define technology priorities as follows: 0 for \ac{4G}, 1 for \ac{5G}, and 2 for \ac{6G}~\cite{LopezPerez2025}.

%%%%%%%%%%%%%%%%%%%%%%%%%%%%%%%%%%%%%%%%%%%%%%%%%%%%%%%%%%%%%%%%%
\vspace{-0.0cm}
\subsection{CSI-RS Beam Codebook and Data Beam Refinement}

Following association, 
beam refinement is assisted by \acp{CSI-RS}, 
which support spatial multiplexing and precise beam alignment. 
In line with the Type-I codebook-based \ac{CSI} feedback framework standardized by \ac{3GPP}, 
each cell transmits a set of precoded \ac{CSI-RS} beams across both polarization panels. 
These beams offer finer angular granularity than \acp{SSB}, 
and are constructed using a 2D-\ac{DFT} codebook that maps antenna geometry into discrete beam directions.

Each \ac{UE} measures the received power of all available \acp{CSI-RS} 
and reports the index of the preferred beam pair (one per polarization). 
This feedback enables the serving cell to select the optimal precoding vectors for downlink transmission. 
Given the two-antenna configuration at the \ac{UE} side, 
this mechanism supports dual-layer \ac{MIMO} and enhances downlink capacity.

The number of CSI-RS beams evaluated in our simulations is as follows:
\begin{itemize}
    \item {4G:} 2, 4, 8, or 64 data beams depending on the number of \acp{TRX} per cell.
    \item {5G:} 64 \ac{CSI-RS} beams per cell, 
    supporting adaptive spatial multiplexing.
    \item {6G:} 128 or 256 \ac{CSI-RS} beams per cell,
    depending on the number of \acp{TRX}, 
    $128$ or $256$, 
    enabling high-resolution multi-user MIMO~\cite{LopezPerez2025}.
\end{itemize}

%%%%%%%%%%%%%%%%%%%%%%%%%%%%%%%%%%%%%%%%%%%%%%%%%%%%%%%%%%%%%%%%%
\vspace{-0.0cm}
\subsection{Traffic-Aware Scheduling Based on CSI-RS Beam Selection}
%David: do we take average counts or a distribution for the number of UEs and PRBs?
%David: Do we use rnadom scheduling or network simplex based? 

The scheduler operates on a data-driven traffic model,
where \ac{PRB} demands are inferred from real-world network measurements. 
Specifically, we leverage hourly deployment statistics---such as the average number of \acp{PRB} utilized per cell and the number of active \acp{UE}---to estimate per-\ac{UE} demand. 
By dividing total \ac{PRB} usage by the number of active \acp{UE}, 
we obtain a reliable estimate of average per-\ac{UE} \ac{PRB} requirement,
which directly guides the resource allocation process.

Each \ac{UE} is equipped with two cross-polarized antennas and configured to receive two spatial streams, 
enabling dual-layer \ac{MIMO} transmission. 
To support this, the scheduler assigns each \ac{UE} a pair of \ac{CSI-RS} beams---one per polarization—selected based on the highest measured \ac{RSRP}, 
following the refinement procedure described earlier.

\Acp{PRB} are allocated independently on a per-beam basis.
Initially, all \acp{PRB}  within each beam are unassigned. 
\acp{UE} are then randomly allocated \acp{PRB} within their selected beam, 
in proportion to their estimated demand, 
until either the demand is fulfilled or the \ac{PRB} pool of the beam is exhausted.
This random allocation strategy is justified by the inherent channel hardening effect of \ac{mMIMO}, 
which significantly reduces per-\ac{PRB} channel variability \cite{dahlman2023}. 

%%%%%%%%%%%%%%%%%%%%%%%%%%%%%%%%%%%%%%%%%%%%%%%%%%%%%%%%%%%%%%%%%
\vspace{-0.0cm}
\subsection{Performance Metrics}

To drive the simulations and evaluate the effectiveness of each deployment strategy,
we assess the following three key performance metrics: 

\subsubsection{SINR and Effective SINR}

The instantaneous \ac{SINR} is computed per \ac{PRB} and per \ac{MIMO} layer,
accounting for transmit power, beamforming gain, path loss, shadowing, fast fading, inter-cell interference, and thermal noise,
in accordance with the adopted \ac{3GPP} \ac{UMa}/\ac{UMi} channel model~\cite{3GPP38901}.

Under the embraced dual-layer \ac{MIMO} scheduling configuration---where each \ac{UE} is assigned two spatial streams---the resulting per-\ac{PRB} and per-layer \acp{SINR} are aggregated into a single effective \ac{SINR} per layer (codeword) using a mutual information-based mapping~\cite{1651855}. 
This effective \acp{SINR} determines the \acp{MCS} that can be reliably supported under prevailing channel and interference conditions.

\subsubsection{Achievable UE Rate} 

The achievable downlink rate for each \ac{UE} is determined by three factors: 
the effective \ac{SINR}, the number of \acp{PRB} allocated via the data-driven scheduling process, and the spatial multiplexing gain provided by dual-layer \ac{MIMO}.
The effective \ac{SINR} governs the \ac{MCS} applied per \ac{MIMO} layer, 
thereby defining the spectral efficiency achieved on each stream.
By combining the spectral efficiency with the number of active layers and allocated \acp{PRB},
we obtain a realistic estimate of per-\ac{UE} throughput, serving as a key indicator of user experience across deployment scenarios.

\subsubsection{Power Usage}

Since different radio technologies exhibit distinct power consumption characteristics,
we adopt a \emph{data-driven, machine learning–based} model to estimate the energy usage of each radio unit across the evaluated deployments.
The model is trained on empirical measurements from commercial multi-technology, multi-band radios and employs an \ac{ANN} to capture the relationship between power draw, hardware configuration, and traffic load~\cite{piovesan2022machine}.

For \ac{4G} and \ac{5G} radios,
all model parameters are directly derived from real-world deployment data.
In contrast, for \ac{6G} radios,
energy usage is estimated by scaling the \ac{5G} model according to forward-looking yet conservative assumptions on component-level efficiency improvements, 
yielding a lower-bound estimate of future \ac{6G} power consumption.

The model also reflects the architectural differences among generations.
The \ac{4G} radios, 
typically co-located across multiple frequency bands, 
may employ \acp{MCPA} to amplify spectrally adjacent carriers using shared amplifier chains,
while widely separated carriers operate on independent power stages.
The non-co-located \ac{5G} radios, by contrast, 
use \acp{AAU} with dedicated amplifier per \ac{TRX},
enabling full digital beamforming at the cost of higher per-chain baseline consumption.
In our \ac{6G} deployments---whether co-located with \ac{5G} or not---each layer is realized through an independent radio unit, 
given the large frequency separation between \ac{5G} and \ac{6G} carriers.

The total power consumption of each radio is decomposed into five key components:
\begin{itemize}
    \item
    \emph{Baseline power consumption:}
    Energy used by always-on circuitry, 
    such as monitoring, synchronization, and control modules.
    
    \item 
    \emph{Baseband processing power:} 
    Computational load required for digital signal processing in \acp{AAU}.
    
    \item 
    \emph{Transceiver power:} 
    Energy drawn by the \acp{TRX}, 
    encompassing digital-to-analog and upconversion.
    
    \item 
    \emph{Power amplifier overhead:} 
    Static bias power required to maintain amplifier readiness.  
    In \ac{4G}, this may correspond to a shared \ac{MCPA} per band, 
    whereas in \ac{5G} and \ac{6G}, 
    each \ac{TRX} integrates its own wideband amplifier.
    
    \item 
    \emph{Radiated output power ($P_{\mathrm{out}}$):} 
     Power needed to generate the transmit signal across the cells served by the radio.  
     This component depends on amplifier and antenna efficiency,
     and scales with the allocated transmission power.
\end{itemize}
\vspace{-0.0cm}
\section{Performance Analysis}
\label{sec:results}

Leveraging our real-world deployment and traffic data from a commercial \ac{4G}/\ac{5G} network in China, 
we conduct a performance analysis of heterogeneous \ac{4G}/\ac{5G}/\ac{6G} networks using \texttt{Giulia}.

%%%%%%%%%%%%%%%%%%%%%%%%%%%%%%%%%%%%%%%%%%%%%%%%%%%%%%%%%%%%%%%%%
\vspace{-0.0cm}
\subsection{Deployment Strategies Evaluated}

The five deployment configurations analyzed are:

\begin{itemize}
    \item 
    \texttt{4G UMa}: 
    Baseline deployment consisting solely of legacy \ac{4G} \ac{LTE} macrocells.
    \item 
    \texttt{4G + 5G UMa}: 
    Empirical dual-layer network with non-co-located \ac{4G} \ac{LTE} and \ac{5G} \ac{NR} macrocells.
    \item 
    \texttt{4G + [5G + 6G UMa] (co-located)}:
    Addition of \ac{6G} macro cells at existing \ac{5G} site locations to enhance bandwidth and spatial multiplexing.
    \item 
    \texttt{4G + 5G + 6G UMi (non-co-located)}: 
    Independent deployment of \ac{6G} microcells targeted at traffic hotspots for localized capacity boosts.
    \item 
    \texttt{4G + 5G + 6G UPi (non-co-located)}:
    Energy-efficient configuration deploying low-power \ac{6G} \ac{UPi} cells at hotspot locations.
\end{itemize}

\begin{figure}[t]
    \centering
    \includegraphics[width=0.4\textwidth]{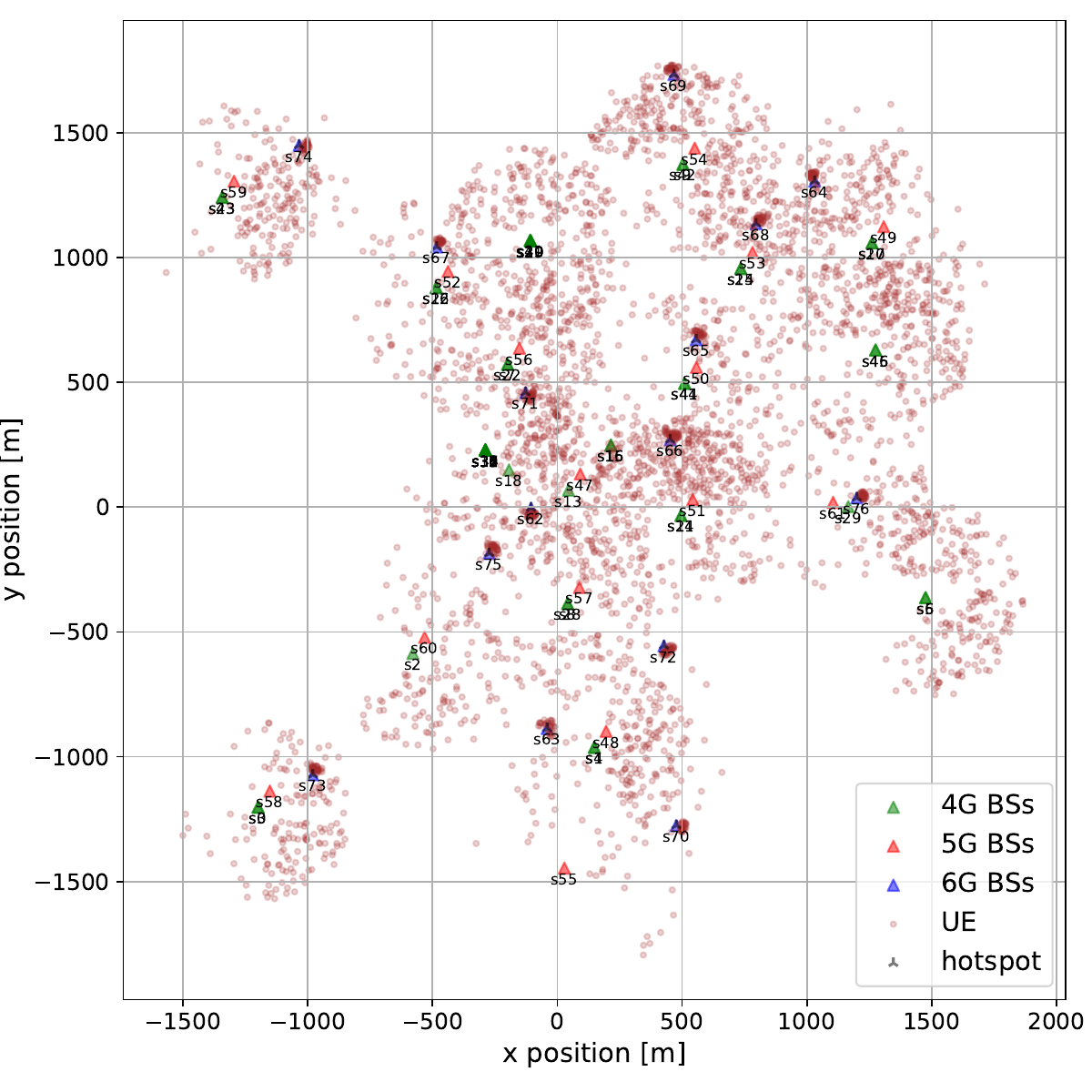}
    \caption{Spatial network layout, along with UE distribution.} % for \texttt{4G\,UMa + 5G\,UMa + 6G\,HS-pico (non-co-located)}
    \label{fig:layout_hspico_ue}
    \vspace{-0.6cm}
\end{figure}

Fig.~\ref{fig:layout_hspico_ue} provides a visual representation of the \texttt{4G + 5G + 6G UPi (non-co-located)} deployment, 
highlighting the spatial distribution of the \ac{4G}, \ac{5G}, and \ac{6G} sites,
and the corresponding \ac{UE} deployment, 
including hotspot locations.
Tab.~\ref{tab:radio_characteristics_data} provides a summary of cells characteristics per technology. 

\begin{table*}[h]
\caption{Data-derived characteristics of 4G and 5G, and those assumed for 6G radios.}
\label{tab:radio_characteristics_data}
{
\centering
\resizebox{1\textwidth}{!}{%
\begin{tabular}{|l||c|c|c|c|c|c|c|c|c|}
\hline
\textbf{Radio} &
\textbf{Carrier freq. [GHz]} &
\textbf{Bandwidth [MHz]} &
\textbf{PRBs} &
\textbf{\# cells} &
\textbf{\acp{TRX} (\(M^{\rm{TRX}}\))} &
\textbf{TX power [dBm]} &
\textbf{Antenna elements} &
\textbf{SSB beams} &
\textbf{CSI-RS beams} \\ \hline\hline

\textbf{4G\,UMa (LTE)} &
1.815--2.664 &
10--20 (mean = 19.1) &
100 &
204 &
2--64 (50\,\%-tile = 4) &
40--52 (50\,\%-tile = 45.9) &
8 (linear)--64 (8$\times$4 dual-pol) &
1 &
2--8 \\ \hline

\textbf{5G\,UMa (NR)} &
2.703 &
100 &
273 &
45 &
64 &
51.3--54.7 (mean 53.2) &
64 (8$\times$4 dual-pol) &
8 &
64 \\ \hline

\textbf{6G\,UMa/UMi} &
10 &
200/400 &
273 &
45 &
128/256 &
55 (58) &
128 (16$\times$4 dual-pol)$^*$ &
16/32 (assumed) &
128/256 (assumed) \\ \hline

\textbf{6G\,UPi} &
10 &
200/400 &
273 &
45 &
128/256 &
52 (55) &
128 (16$\times$4 dual-pol)$^*$ &
16/32 (assumed) &
128/256 (assumed) \\ \hline

\end{tabular}%
}
}
\vspace{0.3em}
{\footnotesize
$^*$ Double in case of 256 TRX with (16$\times$4 dual-pol) and (32$\times$4 dual-pol).
}

%\caption{Data-derived characteristics of 4G and 5G, and those assumed for 6G radios. \\
%$^*$ Double in case of 256 TRX with (16$\times$4 dual-pol) and (32$\times$4 dual-pol).}
% \label{tab:radio_characteristics_data}
\vspace{-0.2cm}
\end{table*}

\vspace{-0.0cm}
\subsection{UE Throughput}

\begin{figure}[b!]
    \centering
    \includegraphics[width=0.45\textwidth]{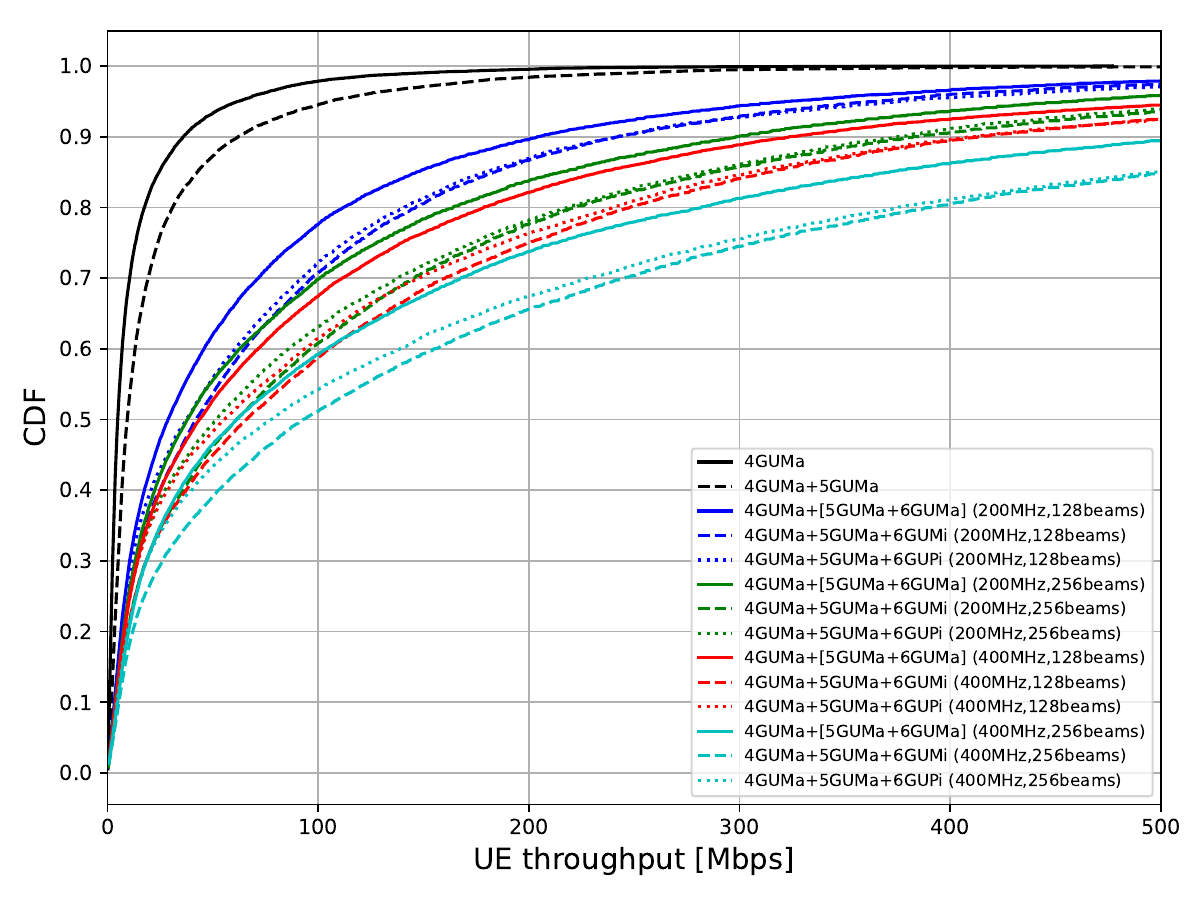}
    \caption{Downlink UE throughput across deployment scenarios.}
    \label{fig:UE_rate_hw}
    \vspace{-0.5cm}
\end{figure}

Fig.~\ref{fig:UE_rate_hw} 
%and~\ref{fig:UE_rate_hw_zoom} 
illustrates the distribution of downlink \ac{UE} rates for all scenarios.
Key statistics—including mean, 5\%-tile, 50\%-tile, and 95\%-tile values—are summarized in Table~\ref{tab:throughput_power_stats}.

\begin{figure}[b!]
    \centering
    \includegraphics[width=0.45\textwidth]{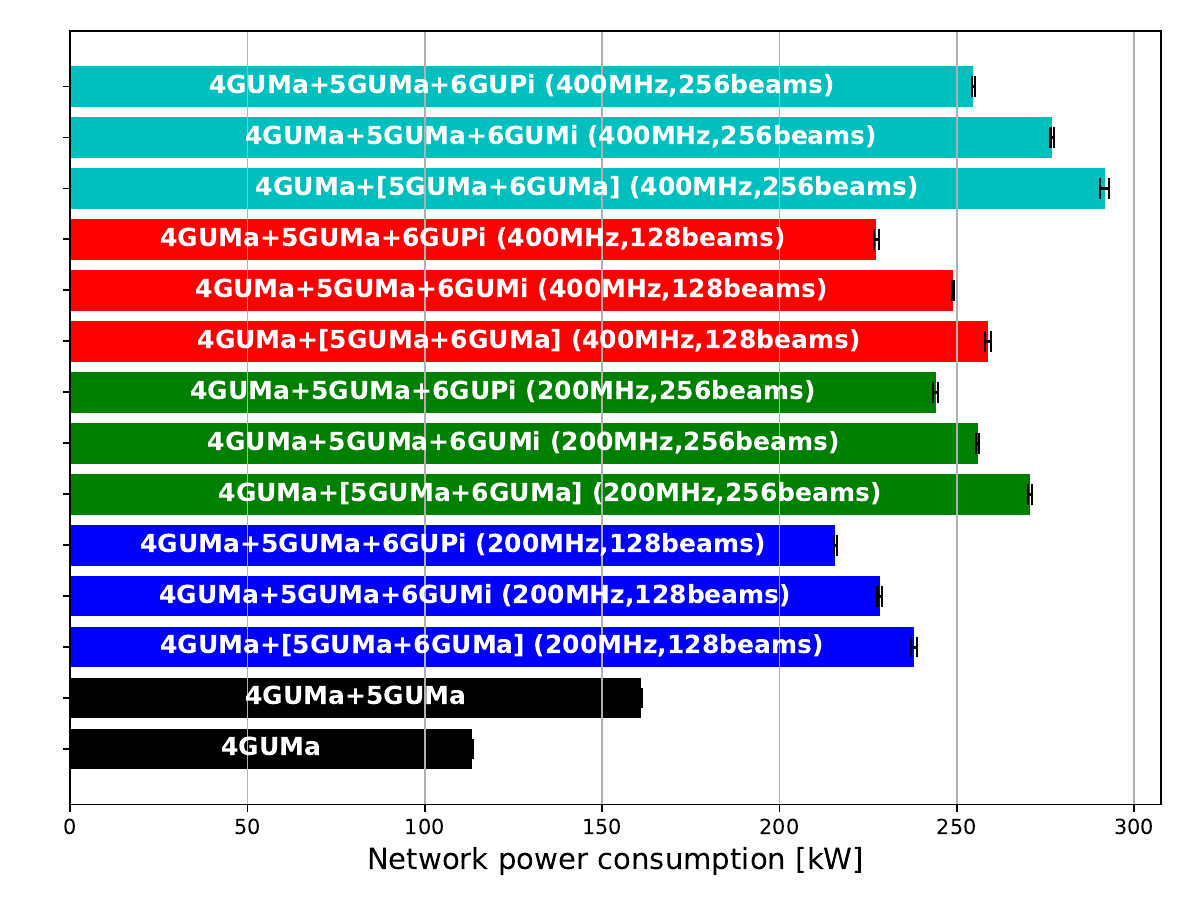}
    \caption{Average network power consumption across scenarios.}
    \label{fig:power_hw}
    \vspace{-0.5cm}
\end{figure}

\begin{comment}
\begin{figure}[t!]
    \centering
    \includegraphics[width=0.4\textwidth]{Figures/UE_throughput_Mbps_hw2.png}
    \caption{Downlink UE throughput (zoom on cell-edge users).}
    \label{fig:UE_rate_hw_zoom}
    \vspace{-0.5cm}
\end{figure}    
\end{comment}

\begin{table*}[ht]
\centering
\caption{Estimated downlink UE throughput statistics derived from the CDF curves
(mean, 5\%, 50\%, and 95\% percentiles). 
%Parentheses indicate gains relative to the
%\texttt{4G\,UMa + 5G\,UMa} baseline; percentage gains are used below 100\%, while
%multiplicative gains are used above 100\%. Values correspond to the data-driven FR3
%cm-wave simulation campaign for City~X (power left blank).
}
\label{tab:throughput_power_stats}
\resizebox{1\textwidth}{!}{%
\begin{tabular}{|l|l||c|c|c|c|c|}
\hline
\textbf{Configuration} & \textbf{Scenario} &
\textbf{Mean [Mbps]} & \textbf{5\%-tile [Mbps]} &
\textbf{ 50\%-tile [Mbps]} & \textbf{95\%-tile [Mbps]} &
\textbf{Power [kW]} \\ \hline\hline

As per data & \texttt{4G\,UMa} &
14.33 (-43\%) &
0.40 (-58\%) &
4.79 (-48\%) &
61.79 (-42\%) & 
112 (-30\%) \\ \hline\hline

As per data & \texttt{4G\,UMa + 5G\,UMa} &
25.34 (1$\times$) &
0.95 (1$\times$) &
9.27 (1$\times$) &
105.55 (1$\times$) &
160 (1$\times$) \\ \hline\hline

\multirow{3}{*}{200MHz/128TRX}
 & \texttt{4G\,UMa + [5G\,UMa + 6G\,UMa]} &
78.93 ($\times$3.1) &
1.62 (+70\%) &
28.70 ($\times$3.1) &
324.31 ($\times$3.1) &
238 (+49\%) \\ \cline{2-7}
 & \texttt{4G\,UMa + 5G\,UMa + 6G\,UMi}  &
95.99 ($\times$3.8) &
1.96 ($\times$2.1) &
41.82 ($\times$4.5) &
365.78 ($\times$3.5) &
228 (+43\%)  \\ \cline{2-7}
 & \texttt{4G\,UMa + 5G\,UMa + 6G\,UPi}  &
94.18 ($\times$3.7) &
1.73 (+81\%) &
37.24 ($\times$4.0) &
376.62 ($\times$3.6) &
215 (+34\%) \\ \hline\hline

\multirow{3}{*}{200MHz/256TRX}
 & \texttt{4G\,UMa + [5G\,UMa + 6G\,UMa]} &
109.38 ($\times$4.3) &
1.76 (+85\%) &
37.82 ($\times$4.1) &
457.53 ($\times$4.3) &
270 (+69\%) \\ \cline{2-7}
 & \texttt{4G\,UMa + 5G\,UMa + 6G\,UMi}  &
142.44 ($\times$5.6) &
2.32 ($\times$2.4) &
61.20 ($\times$6.6) &
573.16 ($\times$5.4) &
255 (+59\%)  \\ \cline{2-7}
 & \texttt{4G\,UMa + 5G\,UMa + 6G\,UPi}  &
135.43 ($\times$5.3) &
1.94 ($\times$2.0) &
51.74 ($\times$5.6) &
544.68 ($\times$5.2) &
245 (+53\%) \\ \hline\hline

\multirow{3}{*}{400MHz/128TRX}
 & \texttt{4G\,UMa + [5G\,UMa + 6G\,UMa]} &
128.86 ($\times$5.1) &
1.89 (+98\%) &
43.67 ($\times$4.7) &
531.42 ($\times$5.0) &
260 (+63\%) \\ \cline{2-7}
 & \texttt{4G\,UMa + 5G\,UMa + 6G\,UMi}  &
161.79 ($\times$6.4) &
2.33 ($\times$2.4) &
65.83 ($\times$7.1) &
638.68 ($\times$6.0) &
250 (+56\%) \\ \cline{2-7}
 & \texttt{4G\,UMa + 5G\,UMa + 6G\,UPi}  &
156.32 ($\times$6.2) &
2.05 ($\times$2.1) &
55.35 ($\times$6.0) &
642.97 ($\times$6.1) &
228 (+43\%)\\ \hline\hline

\multirow{3}{*}{400MHz/256TRX}
 & \texttt{4G\,UMa + [5G\,UMa + 6G\,UMa]} &
192.59 ($\times$7.6) &
2.42 ($\times$2.5) &
61.27 ($\times$6.6) &
829.62 ($\times$7.9) &
292 (+83\%)\\ \cline{2-7}
 & \texttt{4G\,UMa + 5G\,UMa + 6G\,UMi}  &
249.89 ($\times$9.9) &
2.68 ($\times$2.8) &
92.88 ($\times$10.0) &
1027.62 ($\times$9.7) &
275 (+72\%)\\ \cline{2-7}
 & \texttt{4G\,UMa + 5G\,UMa + 6G\,UPi}  &
240.34 ($\times$9.5) &
2.35 ($\times$2.5) &
77.84 ($\times$8.4) &
1014.80 ($\times$9.6) &
255 (+59\%)\\ \hline

\end{tabular}%
\vspace{-0.6cm}
}
\vspace{-0.6cm}
\end{table*}

The baseline \texttt{4G\,UMa} delivers a mean \ac{UE} throughput of 14.33\,Mbps and a 5\,\%-tile of 0.40\,Mbps, 
reflecting 20\,MHz spectrum and simple antenna arrays.
Adding a \ac{5G} layer (\texttt{4G\,UMa + 5G\,UMa}) lifts the mean by +77\,\% and the 5\,\%-tile by 2.4$\times$,
due to 100\,MHz bandwidth and 64-\ac{TRX} \ac{mMIMO}.

Introducing a \ac{6G} layer operating in \ac{FR3} yields substantial further gains.
With 200\,MHz bandwidth and 128 beams,
co-located \ac{6G} macros (\texttt{4G\,UMa + [5G\,UMa + 6G\,UMa]})
already triple the mean and raise the 5\,\%-tile by 70\,\%
relative to the \texttt{4G\,UMa + 5G\,UMa} baseline.
Yet the most significant improvements occur when \ac{6G} nodes are deployed closer to \acp{UE}.
Non-co-located 6G micro (\texttt{6G\,UMi}) and pico/hotspot (\texttt{6G\,UPi}) deployments reduce path loss and improve link geometry,
pushing the 5\,\%-tile throughput up by 1.81--2.1$\times$
and the mean by 3.7--3.8$\times$ compared with \texttt{4G\,UMa + 5G\,UMa}.
The \texttt{6G\,UMi} provides better performance than \texttt{6G\,UPi} due to the larger transmit power. 
This confirms that proximity and spatial densification,
rather than merely co-locating new layers,
are key to unlocking \ac{FR3} capacity.

Considering a 200\,MHz bandwidth, 
increasing the beam count from 128 to 256 in the non-co-located 6G micro-cell (\texttt{6G\,UMi}) deployment improves the 5\,\%-tile by approximately 18\,\% and the mean by about 48\,\% (corresponding to 2.4$\times$ and 5.6$\times$ w.r.t. \texttt{4G\,UMa + 5G\,UMa})
thanks to finer spatial reuse and narrower beam footprints.
However, expanding bandwidth to 400\,MHz, 
while keeping 128 beams, 
provides comparable or slightly higher aggregate gains:
the 5\,\%-tile stays the same w.r.t 200\,MHz bandwidth and 256 beams,
the mean improves another 13\,\%, 
while the 95\,\%-tile rises by 11\,\%.
These results highlight a practical trade-off between spatial multiplexing and bandwidth expansion that operators must carefully evaluate.

Among all cases, 
the 400\,MHz, 256-beam, non-co-located \texttt{6G\,UMi} is best overall,
delivering $\approx$9.9$\times$ mean, $\approx$2.8$\times$ 5\,\%-tile, $\approx$10$\times$ 50\,\%-tile, and $\approx$9.7$\times$ 95\,\%-tile gains relative to \texttt{4G\,UMa + 5G\,UMa}. 
These improvements highlight that closer, data-driven \ac{6G} deployments in \ac{FR3} bands
dramatically enhance both average capacity and user-experience uniformity,
demonstrating that proximity-aware densification---not mere co-location---is the decisive factor in realizing \ac{FR3}’s full potential.

%%%%%%%%%%%%%%%%%%%%%%%%%%%%%%%%%%%%%%%%%%%%%%%%%%%%%%%%%%%%%%%%%
\vspace{-0.0cm}
\subsection{Network Power Consumption}

Figure~\ref{fig:power_hw} illustrates the average total power consumption of the network for each deployment strategy. 
These results highlight the energy trade-offs associated with densification and new-layer integration.

The \texttt{4G\,UMa} baseline consumes 112\,kW, 
i.e., $-30\,\%$ relative to the \texttt{4G\,UMa + 5G\,UMa} benchmark at 160\,kW, 
reflecting the additional bandwidth and activation of 64-\ac{TRX} \ac{mMIMO}.
Introducing \ac{6G} layers in \ac{FR3} increases absolute network power due to wider bandwidths and a larger number of active transceiver chains. 
In our model, each \ac{TRX} contributes approximately 1.5--3\,W, 
leading to near-linear power scaling when the TRX count increases (e.g., 128$\rightarrow$256).

At 200\,MHz/128TRX,
the co-located \texttt{4G\,UMa + [5G\,UMa + 6G\,UMa]} configuration reaches 238\,kW (+$49\,\%$ vs. \texttt{4G\,UMa + 5G\,UMa}), 
whereas closer-to-the-\ac{UE} non-co-located \texttt{6G\,UMi} and \texttt{6G\,UPi} deployments require only 228\,kW (+$43\,\%$) and 215\,kW (+$34\,\%$), respectively,
while delivering substantially higher throughput, as shown earlier. 
This indicates that improved link geometry reduces the required transmit power, limiting the growth of network power despite densification.

More generally, the results highlight a clear throughput--energy trade-off: 
capacity gains come at the cost of higher power,
but the increase is sub-linear relative to throughput.
Notably, the 400\,MHz/128TRX operating point is slightly less power-hungry than 200\,MHz/256TRX at comparable throughput, 
because doubling the number of transceiver chains introduces a hardware power penalty that can outweigh the additional bandwidth-dependent processing (and the associated transmit-power setting in the 400\,MHz mode).

Overall, while \ac{FR3} evolution inevitably raises absolute network power consumption, proximity-aware densification and careful bandwidth/TRX dimensioning allow operators to achieve large throughput gains with a moderate and controlled energy increase, yielding a more favorable power--throughput operating regime than co-located macro-layer upgrades.

\section{Conclusion}
\label{sec:conclusions}

This work evaluated the performance and energy implications of multi-layer \ac{6G} evolution using a data-driven model grounded in a live urban \ac{4G}/\ac{5G} deployment and an \ac{FR3}-based \ac{6G} extension. 
The results show that \ac{FR3} can unlock order-of-magnitude capacity gains,
but only when new layers are deployed based on traffic and spatial data rather than convenience-driven co-location.
In particular, a non-co-located \texttt{6G\,UPi/HS} hotspot deployment achieves approximately 9.5× higher mean \ac{UE} throughput while increasing total network power by only 59\,\% relative to the \texttt{4G\,UMa + 5G\,UMa} baseline. 
This demonstrates that substantial capacity improvements can be obtained with a controlled energy cost when \ac{6G} nodes are placed close to demand hotspots.
Overall, the findings indicate that realizing the full potential of \ac{FR3} is not merely a question of spectrum availability or radio hardware, 
but a deployment optimization problem. Data-driven, proximity-aware densification is essential to maximize throughput gains per unit of energy and to enable practical, scalable, and sustainable \ac{6G} networks.

\begin{comment}
This study evaluated the performance and energy implications of multi-layer \ac{6G} networks by leveraging real-world \ac{4G}/\ac{5G} deployment data from a dense urban area in China and augmenting it with a synthetically constructed \ac{FR3}-based \ac{6G} layer.
By enhancing \texttt{Giulia} with deployment-calibrated parameters,
we systematically explored the trade-offs between co-located and hotspot-targeted \ac{6G} extension strategies.

Our findings are clear:
while \ac{6G} has the potential to significantly boost network capacity,
its energy footprint is substantial---adding a \ac{6G} layer can increase power consumption by up to 35\,\% over current \ac{4G}+\ac{5G} configurations.

Importantly, we observe that convenience-driven co-location of \ac{6G} with existing \ac{5G} sites often fails to capitalize on the spatial reuse and high-capacity benefits of \ac{cm}-wave frequencies,
while still incurring most of the energy cost. 
In contrast, targeted deployment of \ac{6G} micro or pico cells at traffic hotspots delivers superior efficiency, 
achieving up to 7.3× median and 3.5× tail throughput improvements.

These results highlight the need for energy-aware planning in \ac{6G} architecture design. 
Legacy assumptions around site reuse and rapid deployment must give way to strategies that balance performance with sustainability.

Future work will extend this analysis to include uplink modeling, emerging \ac{FR3} massive \ac{MIMO} architectures, and \ac{AI}-driven deployment optimization frameworks.
\end{comment}

%--- Acronyms
\begin{acronym}[AAAAAAAAA]
    \acro{ACLR}{adjacent channel leakage ratio}
    \acro{OFDM}{orthogonal frequency-division multiplexing}
    \acro{PAPR}{peak-to-average power ratio}
    \acro{ANN}{artificial neural network}
    \acro{EESS}{Earth exploration-satellite service}
    \acro{2D-DFT}{two dimensional discrete Fourier transform}
    \acro{3GPP}{3rd Generation Partnership Project}
    \acro{4G}{fourth generation}
    \acro{5G}{fifth generation}
    \acro{6G}{sixth generation}
    \acro{AAU}{active antenna unit}
    \acro{AF}{array factor}
    \acro{AH}{aerial highway}
    \acro{AI}{artificial intelligence}
    \acro{AoA}{angle of arrival}
    \acro{AoD}{angle of departure}
    \acro{BBU}{baseband unit}
    \acro{BO}{bayesian optimization}
    \acro{BS}{base station}
    \acro{BVLoS}{beyond visual line of sight}
    \acro{CAGR}{compound annual growth rate}
    \acro{CCUAV}{cellular connected unmanned aerial vehicle}
    \acro{CDF}{cumulative distribution function}
    \acro{cm-wave}{centimeter wave}
    \acro{CQI}{channel quality indicator}
    \acro{CRS}{common reference signal}
    \acro{CSI}{channel state information}
    \acro{CSI-RS}{channel state information-reference signal}
    \acro{D2D}{device to device}
    \acro{DFT}{discrete Fourier transform}
    \acro{DL}{downlink}
    \acro{DoF}{degree of freedom}
    \acro{eGA}{elite genetic algorithm}
    \acro{eICIC}{enhanced inter-cell interference coordination}
    \acro{E}{eastern}
    \acro{ES}{Eigenscore}
    \acro{FR1}{frequency range 1}
    \acro{FR2}{frequency range 2}
    \acro{FR3}{frequency range 3}
    \acro{GA}{genetic algorithm}
    \acro{gUE}{ground user equipment}
    \acro{HO}{handover}
    \acro{ICC}{international conference on communications}
    \acro{IMT}{international mobile telecommunication system}
    \acro{ISD}{inter-site distance}
    \acro{IUD}{inter-UAV distance}
    \acro{ITU}{International Telecommunication Union}
    \acro{KPI}{key performance indicator}
    \acro{LEO}{low Earth orbit}
    \acro{LoS}{line of sight}
    \acro{LTE}{long term evolution}
    \acro{MAMA}{mMIMO-Aerial-Metric-Association}
    \acro{MCPA}{multicarrier power amplifier}
    \acro{MCS}{modulation and coding scheme}
    \acro{MINP}{mixed-integer nonlinear problem}
    \acro{ML}{machine learning}
    \acro{MIMO}{multiple-input multiple-output}
    \acro{mMIMO}{massive multiple-input multiple-output}
    \acro{mmWave}{millimeter wave}
    \acro{MNO}{mobile network operator}
    \acro{MU-mMIMO}{multi-user massive multiple-input multiple-output}
    \acro{MU-MIMO}{multi-user multiple-input multiple-output}
    \acro{NLoS}{non line of sight}
    \acro{NOMA}{non-orthogonal multiple access}
    \acro{NR}{new radio}
    \acro{PBCH}{physical broadcast channel}
    \acro{P2P}{point to point}
    \acro{PAHSS}{Particle Aerial Highway Swarm Segmentation}
    \acro{PL}{path loss}
    \acro{PMI}{precoding matrix indicator}
    \acro{PRB}{physical resource block}
    \acro{PSO}{particle swarm optimization}
    \acro{PSS}{primary synchronization signal}
    \acro{QoS}{quality of services}
    \acro{RAN}{radio access network}
    \acro{RE}{resource element}
    \acro{RI}{rank indicator}
    \acro{RRC}{radio resource control}
    \acro{RSRP}{reference signal received power}
    \acro{RSS}{received signal strength}
    \acro{SRS}{sounding reference signal}
    \acro{SSB}{synchronization signal block}
    \acro{SINR}{signal-to-interference-plus-noise ratio}
    \acro{SO}{southern}
    \acro{SSS}{secondary synchronization signal}
    \acro{SVD}{single value decomposition}
    \acro{thp}{throughput}
    \acro{TRX}{transceiver}
    \acro{UAM}{urban air mobility}
    \acro{UAV}{unmanned aerial vehicle}
    \acro{UE}{user equipment}
    \acro{UL}{uplink}
    \acro{UMa}{urban macro}
    \acro{UMi}{urban micro}
    \acro{UPA}{uniform planar array}
    \acro{UPi}{urban pico}
    \acro{URD}{Urban Random Distributed}
    \acro{URLLC}{ultra-reliable low latency communication}
    \acro{WRC}{world radiocommunication conference}
    \acro{ZF}{zero forcing}
    \acro{mm}{millimeter}
    \acro{cm}{centimeter}
    \acro{RF}{radio frequency}
    \acro{CMOS}{complementary metal--oxide--semiconductor}
    \acro{SiGe}{silicon--germanium}
    \acro{GaN}{gallium nitride}
    
\end{acronym}

%\newpage
\bibliographystyle{IEEEtran}
\bibliography{journalAbbreviations, main}

\end{document}